\documentclass[10pt]{article}
\usepackage[letterpaper]{geometry}
\usepackage{hicss}
\usepackage{times}
\usepackage[none]{hyphenat}
\usepackage{url}
\usepackage{latexsym}
\usepackage{minted}
\usepackage{indentfirst}
\usepackage{graphicx}
\graphicspath{{images/}}
\usepackage[
    style=apa,
  ]{biblatex}
\usepackage{booktabs}
\usepackage{array}     
\usepackage{graphicx}
\usepackage{listings}
\usepackage[table]{xcolor}

\definecolor{rowlight}{gray}{0.93}
\definecolor{rowdark}{gray}{0.82}
\usepackage{times}
\usepackage{balance}

\usepackage{booktabs}        % \toprule \midrule \bottomrule
\usepackage{array}           % p{} column type
\usepackage{multirow}        % \multirow if needed later

\usepackage[table]{xcolor}   % row colors in tables + general color

\usepackage{tcolorbox}       % colored boxes
\usepackage{amsmath}         % equation environments
\tcbuselibrary{skins, breakable}  % tcolorbox libraries

\usepackage{graphicx}        % \includegraphics
\usepackage{tikz}            % if we draw pipeline figure in tikz
\usetikzlibrary{arrows.meta, shapes, positioning, fit, backgrounds}
\usepackage{textcomp}
\usepackage{hyperref}        % clickable refs — check HICSS allows this
\usepackage[T1]{fontenc}
\usepackage{cleveref}        % \cref instead of \ref — optional

\definecolor{rowlight}{gray}{0.93}
\definecolor{rowdark}{gray}{0.82}

\tcbset{
    formulabox/.style={
        enhanced,
        breakable,
        float,
        colback=gray!8,
        colframe=gray!60,
        fonttitle=\bfseries\small,
        coltitle=white,
        attach boxed title to top left={
            yshift=-2mm, xshift=4mm},
        boxed title style={
            colback=gray!60, rounded corners},
        rounded corners,
        boxrule=0.6pt,
        left=4pt, right=4pt,
        top=6pt, bottom=4pt,
        fontupper=\small
    }
}
\usetikzlibrary{arrows.meta, shapes.geometric, positioning, 
                fit, backgrounds, shadows.blur}
                
\lstdefinestyle{jsonbox}{
    basicstyle=\ttfamily\tiny,
    breaklines=true,
    breakatwhitespace=false,
    breakindent=0pt,
    postbreak=\mbox{\textcolor{gray}{$\hookrightarrow$}\space},
    columns=fullflexible,
    keepspaces=true,
    upquote=true,
    frame=single,
    rulecolor=\color{gray!60},
    backgroundcolor=\color{gray!8},
    aboveskip=4pt,
    belowskip=4pt,
    xleftmargin=3pt,
    xrightmargin=3pt
}
\title{From Code to Requirements: Agentic Reverse Engineering of Business Rules at Enterprise Scale}

\author{Garima Agrawal \quad Prasun Das \quad Priyanka L \quad Minisha N \\
 Akshay Sarvade \quad Hemath Manivanan \quad Sravani Joshna \quad Prasad Kalyansundaram \\[6pt]
 {\large Cognizant, Plano, Texas, USA}\\[6pt]
 {\{garima.agrawal, prasun.das, priyanka.lnu2307c7, minisha.n, akshay.sarvade,} \\
 {hemathKumar.manivannan, malla.sravaniJoshna, prasad.kalyanasundaram\}@cognizant.com}}

\begin{document}
\maketitle
% \begingroup
% \renewcommand{\thefootnote}{\fnsymbol{footnote}}
% \setcounter{footnote}{1}
%  \footnotetext{Submitted to HICSS-60.}
% \setcounter{footnote}{0}
% \endgroup

\begin{abstract}
Business requirements for enterprise software systems are rarely 
captured in structured form; the logic resides instead in source 
code, configuration files, and institutional memory. When these 
systems must be migrated, extended, or audited, the absence of 
formal requirements artifacts forces teams into expensive, 
knowledge-losing manual reverse engineering. This paper presents an 
agentic framework that autonomously generates Business Requirements 
Documents (bRDs) through reverse engineering of undocumented 
enterprise software. Seven specialized agents collaborate to 
discover user journeys, extract business rules, and synthesize a 
locked bRD from code, test suites, configuration, and available 
documentation, supported by static analysis tools, guided by 
embedded expert practitioner cognitive models, and overseen by human 
reviewers at controlled escalation points. Measured on actual 
execution traces across a real enterprise deployment, the framework 
generates comprehensive bRDs in under nine minutes per service, with 
extracted rules independently corroborated against real production 
defect records. In a comparable prior migration to the same target 
architecture, delivery required over two years; on the present 
programme the framework achieves a cost reduction exceeding 98\% 
against a baseline derived from actual repository metrics using 
IFPUG complexity models and industry benchmark labor rates.
\end{abstract}
\subsection*{Keywords:} Agentic AI, Requirements Reverse Engineering, Legacy systems, SDLC automation, Multi-Agent Systems

\section{Introduction}
\label{sec:intro}

Enterprise software systems accumulate business logic over years of 
evolution, encoding rules and behaviors directly into source code, 
configuration files, and the institutional memory of personnel who 
may no longer be available. Modern microservice architectures face 
the same documentation gap through different mechanisms, including 
sprint velocity pressures, shifting team ownership, and the implicit 
assumption that working code is sufficient specification. In both 
cases, the Business Requirements Document (bRD), which formally 
captures user journeys, business rules, acceptance criteria, gap 
analyses, and risk dependencies, is absent precisely when it is 
most needed.

The consequences are well understood in practice. Migration programs 
stall on undiscovered behavioral edge cases. Compliance audits 
require expensive manual archaeology. Producing a bRD manually 
demands a cross-functional team of business analysts, developers, 
quality engineers, and domain experts working in concert for weeks 
per component, decoding behavior from source code, test suites, 
configuration files, and tribal knowledge. Knowledge earned on one 
component does not transfer to the next. Every engagement restarts 
from near-zero.

Recent advances in LLM-based agentic systems have addressed adjacent 
problems, including code generation~\cite{qian2024}, automated 
testing~\cite{schafer2024}, and architectural 
recovery~\cite{canfora2007}, but requirements reverse engineering 
from undocumented enterprise systems remains largely unaddressed. 
Where approaches exist, three critical limitations persist. First, 
existing tools rely on single data sources, typically source code 
alone, missing the distributed behavioral signal encoded across 
repositories, test suites, configuration files, and documentation 
simultaneously. Second, current agentic approaches exhibit heavy 
LLM dependence, invoking generative models for tasks better handled 
by deterministic static analysis such as AST parsing, pattern 
detection, and schema diffing, resulting in unnecessary token 
consumption and compounding hallucination risk. Third, no existing 
framework encodes domain expertise as a first-class architectural 
element. A senior business analyst questions differently than a 
retrieval agent; a quality engineer reads a test suite differently 
than a code parser. The cognitive models of experienced 
practitioners, including how they hunt edge cases, detect ambiguity, 
and surface regulatory risk, are absent from current 
designs~\cite{hong2024}.

A single-prompt approach cannot meet this challenge at enterprise 
scale. A component spans tens of thousands of lines across multiple 
repositories, together with tests, configuration, and documentation, 
exceeding any single context window; a model asked to hold all of it 
fills the gaps with plausible invention. The required reasoning also 
differs by expertise, from journey analysis to test-suite reading to 
regulatory scrutiny, and no single prompt performs all of these well. 
This paper presents a seven-agent cognitive framework that decomposes 
the work across specialized agents, each with a narrow brief, a 
deterministic tool belt, and a defined output contract. It addresses 
the three limitations directly: it integrates evidence from code, 
test suites, configuration, and documentation simultaneously; it 
employs static analysis as the primary extraction mechanism, 
reserving LLM reasoning for synthesis, conflict resolution, and 
plain-English translation; and it encodes expert practitioner 
cognitive models as agent-level capabilities that shape how each 
agent reasons, not merely what it retrieves. A knowledge layer 
enables reuse across successive runs.

Critically, this is not a research prototype. The framework has been 
deployed on a large-scale enterprise environment spanning modern 
microservices and legacy systems, under human supervision: agents 
execute autonomously but escalate to human reviewers at defined 
decision points. The magnitude of the problem is concrete, as a 
comparable prior migration to the same target architecture for 
another product line required over two years of effort in practice. 
Against a baseline derived from actual repository metrics, manual bRD 
generation for a single service requires 117 person-hours on average, 
whereas the framework generates an equivalent bRD in a generation 
time of under nine minutes at an average LLM inference cost of \$0.07 
per service, both measured from actual execution traces. This 
generation time is distinct from human review and validation effort, 
which we report separately in Section~\ref{sec:implementation}.

The contributions of this paper are as follows:
\begin{enumerate}
    \item A seven-agent cognitive architecture for requirements 
    reverse engineering that integrates multi-source evidence with 
    embedded domain expertise.
    \item A knowledge layer that supports reuse across successive 
    runs and incremental regeneration as components change.
    \item An effort and cost evaluation grounded in actual repository 
    metrics and measured execution traces, using IFPUG-derived 
    complexity models, comprehension-based effort modeling, and 
    industry benchmark labor rates.
    \item Artifact quality validation in which extracted rules are 
    corroborated against independent production defect records, 
    across both modern and legacy systems.
\end{enumerate}

The remainder of this paper is organized as follows. 
Section~\ref{sec:related} reviews related work. 
Section~\ref{sec:framework} describes the framework architecture. 
Section~\ref{sec:implementation} details the implementation and 
evaluation. Section~\ref{sec:discussion} discusses the results and 
future work. Section~\ref{sec:conclusion} concludes.

\section{Related Work}
\label{sec:related}

The three limitations identified in Section~\ref{sec:intro} 
position this work within three research threads: requirements 
recovery, multi-agent LLM systems, and software effort 
estimation. \textbf{Requirements Recovery and Reverse Engineering -}
Chikofsky and Cross~\cite{chikofsky1990} established the 
foundational taxonomy of reverse engineering, defining 
requirements recovery as its highest abstraction level. Canfora 
and Di Penta~\cite{canfora2007} note that most approaches recover 
structural artifacts such as call graphs and class diagrams 
rather than behavioral requirements, operate on single source 
types, and do not address the synthesis problem inherent in 
enterprise systems where behavioral signal is distributed across 
code, tests, configuration, and documentation. The recovery of 
user journeys, business rules, and acceptance criteria from 
undocumented systems remains largely unaddressed. \textbf{Multi-Agent LLM Systems for Software 
Engineering -} ChatDev~\cite{qian2024} frames development as a 
virtual organization of communicating agents; 
MetaGPT~\cite{hong2024} encodes standardized operating procedures 
into role-based workflows; and SWE-agent~\cite{yang2024} resolves 
software issues through iterative code analysis. AgentModernize~\cite{ahmed2026agentmodernize} addresses the 
downstream conversion of legacy code to modern components, reporting 
91.2\% business rule capture via a Behavioral Specification Graph; 
however, it operates on the modernization step and does not encode 
the questioning behavior of business, quality, and compliance 
specialists that our framework embeds as agent-level cognitive 
models. Related industrial work has demonstrated 
agentic pipelines for enterprise knowledge 
extraction~\cite{agrawal2024beyondrag} and knowledge-graph 
augmentation to reduce hallucination risk~\cite{agrawal2024kg}. 
All operate downstream of requirements, assuming specifications 
exist. Our framework addresses the upstream problem they 
presuppose: recovering behavioral specifications from systems 
where none were ever written. \textbf{Effort Estimation in Software Engineering -}
COCOMO~II~\cite{boehm2000} provides the foundational parametric 
effort model, and IFPUG Function Point Analysis~\cite{ifpug2009} 
a complexity measure grounded in functional components with 
industry-derived weights. ISBSG benchmark data~\cite{isbsg2017} 
supplies the Specify and Design phase ratios used in our manual 
baseline, while program comprehension 
studies~\cite{minelli2015,xia2018} establish that 52--70\% of 
maintenance effort is consumed by code reading, anchoring our 
comprehension rate of 100~LOC/hour. To our knowledge, no prior 
work has applied IFPUG-derived complexity weights specifically to 
reverse engineering effort estimation, an additional 
methodological contribution of this work.

\section{Framework Design}
\label{sec:framework}

\begin{figure*}[t]
\centering
\includegraphics[width=\textwidth]{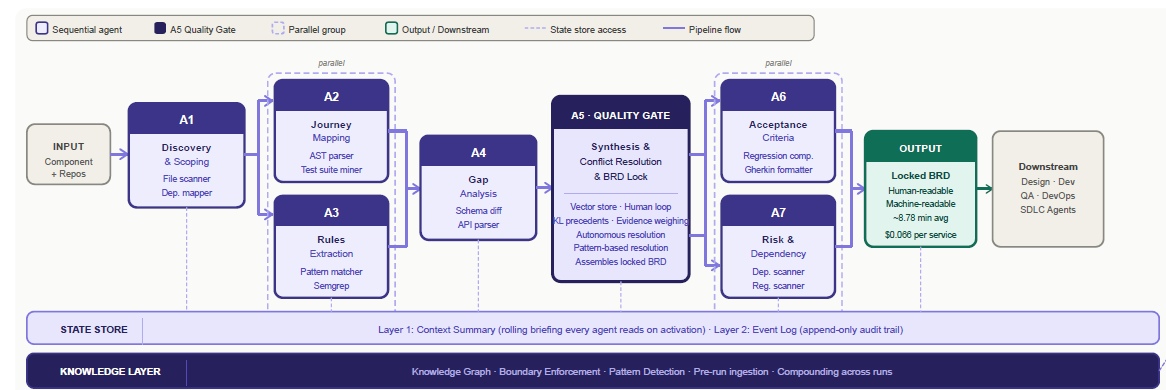}
\caption{Seven-agent bRD generation pipeline. Agents~2+3 and 
Agents~6+7 execute in parallel. All agents read from and write 
to the shared state store. The knowledge layer compounds 
organizational knowledge across successive pipeline runs.}
\label{fig:pipeline}
\end{figure*}

The gap established in Section~\ref{sec:related}, that existing 
approaches neither integrate behavioral signal across multiple 
source types nor encode domain expertise architecturally, 
motivates the design presented here. Rather than a monolithic 
prompt-driven system, we designed a cognitive pipeline of 
specialized agents that collaborate, share structured state, and 
defer to static analysis wherever deterministic extraction is 
possible, reserving LLM reasoning for the points where 
human-level synthesis is genuinely required. 
Figure~\ref{fig:pipeline} illustrates the end-to-end pipeline.

\subsection{System Overview}

The pipeline accepts a software component as input, defined by its 
repository set together with whatever supporting artifacts exist: 
source code accessed through a Git Nexus connector, Jira issues, Confluence 
pages accessed through an MCP connector, API specifications, runtime 
call stacks, and application logs. These sources are read and parsed 
directly in full. The pipeline produces a locked bRD in two forms: a 
human-readable document for business and technical stakeholders, and 
a machine-readable json artifact for downstream SDLC agents in design, 
development, QA, and DevOps. A LangGraph orchestrator~\cite{langgraph} 
manages sequencing, parallel execution, confidence monitoring, and 
knowledge layer graduation. Agent~1 establishes scope; Agents~2 
and~3 then run in parallel, both reading only the scope definition 
from Agent~1 and producing independent outputs: Agent~2 a journey 
inventory, Agent~3 a rules catalogue. Because they run concurrently, 
neither depends on the other's output; the authoritative 
cross-linking of rules to journeys is performed later by Agent~5, 
which cross-examines all upstream outputs, resolves conflicts, and 
can dispatch corrections back to the responsible agent before 
locking. Agent~4 runs once both complete; Agent~5 is the quality 
gate that runs when all prior outputs are available; and Agents~6 
and~7 run in parallel after the bRD is locked. A shared state store 
threads through the pipeline, carrying context forward and preserving 
a full audit trail. No prior bRD or knowledge base is assumed; this 
cold-start setting is the one this paper evaluates, and the knowledge 
layer (Section~\ref{sec:framework}) describes how prior knowledge is 
reused when it does exist.

\subsection{Agent DNA and Cognitive Model}

Every agent carries a shared base schema, termed Agent DNA, 
comprising six blocks: identity, skills, domain knowledge, 
cognitive model, house rules, and pipeline position. Three of 
these blocks encode direct responses to the three limitations 
of Section~\ref{sec:intro}.

\textit{Multi-source integration} is encoded in the domain 
knowledge block: each agent knows where behavioral signal lives 
across production code, test suites, configuration, documentation, 
and issue trackers, and must consult a minimum of two independent 
sources before accepting any finding as valid.

\textit{Static-first extraction} is encoded in the tools block 
and house rules: each agent queries a registry of pre-built static 
analysis capabilities, performing AST parsing, pattern detection, 
schema diffing, and test suite mining deterministically. The LLM 
is invoked only for translation, conflict resolution, and 
narrative synthesis, operating on curated structured inputs rather 
than raw repository content, which minimizes token consumption and 
hallucination risk.

\textit{Embedded domain expertise} is encoded in the cognitive 
model block: each agent carries the thinking modes of a senior 
business analyst, including systems thinking, skeptical reading 
that treats documentation as hypothesis, edge case hunting, and 
ambiguity detection. These are operationalized through 
non-negotiable house rules: scope is never silently expanded, 
conflicts never silently resolved, confidence never inflated, and 
every assumption named, tracked, and owned.

\subsection{The Seven Agents}

\textbf{Agent~1: Discovery and Scoping.} The component under 
analysis is defined by a human expert, not by the pipeline: an 
engineer maps a named service to its candidate repositories and 
configures the run with the supporting sources listed above. Taking 
this human-provided scope as its starting point, Agent~1 scans 
repository file trees, catalogs source, configuration, test, and 
documentation artifacts, scores each for freshness and authority, 
maps upstream and downstream dependencies, and produces the scope 
definition all subsequent agents build on. It confirms which 
repositories, endpoints, and dependencies genuinely belong to the 
component and flags any that appear out of scope. Ambiguous 
boundaries trigger a human decision request before the pipeline 
proceeds.

\textbf{Agent~2: Journey Mapping} enumerates every user and system 
journey through the component. It operates across four source types 
simultaneously: API endpoints and UI routes located by static 
scanning; code logic traced by AST parsing; documentation parsed 
by the document adapter; and test suites mined for encoded 
behavioral specifications. Test suite mining is architecturally 
significant: regression tests encode the journeys that development 
teams actually validated, including edge cases and error paths never 
formally documented. Actors and roles are extracted from 
authentication and authorization checks, endpoint access controls, 
and role references in tests and configuration, rather than assumed. 
Agent~2 produces a typed journey inventory covering happy paths, 
alternate paths, error paths, and edge cases per actor and channel, 
together with a test-to-journey coverage matrix.

\textbf{Agent~3: Business Rules Extraction} surfaces every business 
rule embedded in source code, configuration files, and 
documentation. It applies pattern-based static analysis, using AST 
queries and Semgrep-style rule patterns over the parsed syntax tree, 
to identify conditional logic, validation guards, eligibility 
checks, and threshold constants, each returned with its file 
location and surrounding context before the LLM translates it into 
a plain-English rule with actor, condition, and outcome explicitly 
stated. Agent~3 produces a preliminary rule-to-journey linking, 
flags rules it cannot yet map, and raises regulatory signals for 
PII, financial, and compliance-relevant logic; the authoritative 
rule-to-journey linking is reconciled by Agent~5 once both 
inventories are available.

\begin{tcolorbox}[formulabox, title=Business rule with provenance (masked)]
\footnotesize\ttfamily
\begin{verbatim}
{ "rule_id": "R-014",
  "statement": "Requests from an
    unsupported region are rejected
    before pricing runs",
  "source": ["EligibilityService.java:118",
             "regions.yaml"],
  "corroborated_by": ["RegionValidatorTest"],
  "confidence": "HIGH",
  "regulatory_flag": false,
  "journeys": ["J-02","J-05"],
  "assumptions": [] }
\end{verbatim}
\end{tcolorbox}

\textbf{Agent~4: Gap Analysis} compares legacy behavior against 
target architecture capabilities for every journey and business rule 
from Agents~2 and~3. It requires a target-state specification as 
input; where a target schema is available, a schema diff compares 
the legacy data model, entity fields and types extracted from code 
and configuration, against the target to identify added, removed, 
renamed, or retyped fields, and where none exists Agent~4 reports 
only the discovered baseline and flags the absence. Gaps are 
classified as clean map (legacy behavior maps directly), transform 
needed (an adaptation layer is required), missing (absent from the 
target), or deferred (resolution planned for a later cycle). Each 
gap is risk-scored by business impact and customer-facing exposure, 
and blocking gaps that prevent bRD lock are escalated to the human 
loop.

\textbf{Agent~5: Synthesis, Conflict Resolution, and bRD Lock} is 
the quality gate of the pipeline. It cross-examines the outputs of 
Agents~1 through~4 simultaneously, detecting conflicts within and 
across agent outputs, coverage holes, confidence chain violations, 
and assumption dependency chains. Resolution proceeds in three 
phases: autonomous resolution using knowledge layer precedents and 
evidence weighing; pattern-based resolution for known conflict 
types, applying fixed precedence rules such as test evidence 
overriding documentation (Section~\ref{sec:framework}, Artifact 
Composition); and human loop escalation only when a conflict 
requires a business judgment that cannot be resolved analytically. 
Human escalation surfaces a structured decision package comprising 
the exact conflict statement, evidence on both sides, a 
recommendation, and a single answerable question. Every human 
decision is logged permanently with full rationale. Agent~5 then 
assembles all validated artifacts into a coherent locked bRD in 
both human-readable and machine-readable form.

\begin{tcolorbox}[formulabox, title=A conflict resolved by precedence (masked)]
\footnotesize\ttfamily
\begin{verbatim}
CONFLICT  C-07
  Journey J-04: an unlocked item is
    always eligible.
  Rule R-018 + test SuiteB#rejectsForeignLock:
    rejected when locked to another provider.
RESOLUTION
  Test evidence prevails; J-04 amended,
  R-018 retained, exception branch added.
  logged: rationale, sources, confidence 0.91
\end{verbatim}
\end{tcolorbox}

\textbf{Agent~6: Acceptance Criteria} translates every locked 
journey and business rule into verifiable Given/When/Then acceptance 
criteria formatted as Gherkin scenarios~\cite{gherkin} for direct 
adoption in Behavior-Driven Development (BDD) test frameworks. It 
compares new criteria against the existing regression suite to 
identify coverage gaps and generates structured new test case 
suggestions.

\textbf{Agent~7: Risk and Dependency} maps all system, data, and 
team dependencies, scans for regulatory and compliance obligations, 
scores each risk by likelihood and business impact, and produces a 
structured risk register fed into the locked bRD\@.
Table~\ref{tab:agents} summarizes the input sources, static tools, 
LLM role, and output artifact for each agent.

\begin{table*}[t]
\centering
\caption{Agent pipeline: input sources, static tools, LLM role, 
and output artifact.}
\label{tab:agents}
{\small
\rowcolors{2}{rowlight}{white}
\begin{tabular}{lp{3.0cm}p{3.0cm}p{2.8cm}p{3.0cm}}
\toprule
\rowcolor{rowdark}
\textbf{Agent} & 
\textbf{Input Sources} & 
\textbf{Static Tools} & 
\textbf{LLM Role} & 
\textbf{Output} \\
\midrule
A1 Discovery & 
Repos, docs, tickets & 
File scanner, dependency mapper & 
Scope synthesis & 
Scope doc, artifact catalog \\
\addlinespace[3pt]
A2 Journey & 
Code, test suites, API specs, docs & 
AST parser, test suite miner & 
Journey typing, conflict resolution & 
Journey inventory, coverage matrix \\
\addlinespace[3pt]
A3 Rules & 
Source code, config files, docs & 
Pattern matcher, AST parser & 
Rule translation & 
Rules catalog, regulatory flags \\
\addlinespace[3pt]
A4 Gap & 
Journey inventory, rules, target specs & 
Schema diff, API parser & 
Gap classification & 
Gap register, risk scores \\
\addlinespace[3pt]
A5 Synthesis & 
All agent outputs, knowledge layer & 
Vector store & 
Conflict resolution, bRD assembly & 
Locked bRD (human + machine) \\
\addlinespace[3pt]
A6 AC & 
Locked bRD, test suite & 
Regression comparator & 
AC authoring & 
AC document, Gherkin feature file \\
\addlinespace[3pt]
A7 Risk & 
Scope, journeys, rules, gaps & 
Dependency scanner & 
Risk narrative & 
Risk register, dependency map \\
\bottomrule
\end{tabular}
}
\end{table*}

\subsection{State Store and Human Loop}

Each run maintains a two-layer state store. Layer~1 is a rolling 
context summary, the briefing every agent reads on activation, 
carrying current scope, summaries, open items, confidence flags, 
and handoff notes from the preceding agent. Layer~2 is an 
append-only immutable event log recording every finding, decision, 
tool call, and conflict with full provenance: agent, timestamp, 
source references, confidence, and graduation flags. The summary 
is working memory; the log is the audit trail.

The human loop is a controlled escalation mechanism, not a 
default. Agent~5 escalates only when a conflict cannot be resolved 
from available evidence, an assumption chain exceeds its risk 
threshold, or confidence on a critical-path journey falls below 
the minimum, surfacing a structured decision package and pausing 
the pipeline pending response.

\subsection{The Knowledge Layer}

The knowledge layer separates this framework from single-run 
approaches. It maintains a knowledge graph whose nodes represent 
the core artifacts of requirements engineering, components, 
journeys, business rules, actors, data entities, gaps, and 
decisions, each carrying full provenance. Relationships are 
classified as structural, discovered, inferred, or emergent, with 
emergent patterns crystallizing when the same finding recurs 
across three or more components. The layer can serve prior 
knowledge when it exists, enforces domain and regulatory boundaries 
on reads and writes, detects conflicts, and quality-gates findings 
before they become permanent.
Access follows the two operating modes. Within a single cold-start 
run, agents persist their structured artifacts to the store and 
later agents read the specific named artifacts they require 
directly, by identity rather than by similarity search. This keyed 
access avoids threading all accumulated state through the 
orchestrator, which would otherwise exceed context limits and lose 
information. Across runs, when a prior knowledge base or partial bRD 
exists, a vector index enables retrieval of previously established 
rules, journeys, and decisions so that incremental changes to a 
component reuse existing knowledge rather than rediscovering it. 
This cross-run reuse and the resulting reduction in token cost are 
architected properties of the knowledge layer; they are not among 
the cold-start results reported in Section~\ref{sec:implementation}.

\subsection{Artifact Composition and Traceability}

The agents do not produce independent outputs stitched together 
at the end; each acceptance criterion is generated from the 
intersection of three upstream artifacts, with each mapping to 
one clause of a Given/When/Then specification:
\vspace{0.1em}
\begin{equation*}
\text{AC} = \underbrace{\text{Rule}}_{\text{Given}} \;+\; 
            \underbrace{\text{Journey step}}_{\text{When}} \;+\; 
            \underbrace{\text{Test / Gap closure}}_{\text{Then}}
\end{equation*}

\noindent A behavior is treated as expected only where all three 
intersect, making every criterion bidirectionally traceable to a 
governing business rule (Agent~3), a journey step (Agent~2), and 
executable test evidence. Where sources conflict, resolution 
follows a fixed precedence:
\vspace{0.1em}
\begin{equation*}
\text{Test evidence} \;\succ\; \text{Business rule} 
\;\succ\; \text{Journey}
\end{equation*}

\noindent Test evidence is treated as ground truth, business rules 
override aspirational journeys, and conflicting journeys are 
realigned to the production baseline. This precedence is what 
allows Agent~5 to lock the bRD deterministically rather than 
through free-form synthesis, ensuring generated criteria remain 
verifiable rather than aspirational.

\section{Implementation and Evaluation}
\label{sec:implementation}

\subsection{System Configuration}

The framework described in Section~\ref{sec:framework} was 
deployed on a real enterprise Java codebase and evaluated across 
two use cases: modern undocumented microservices and a legacy 
migration pilot. The pipeline is orchestrated using 
LangGraph~\cite{langgraph} with LangSmith~\cite{langsmith} 
providing agent observability, token tracing, and execution 
monitoring across all runs. Repository access is provided through 
the Model Context Protocol (MCP) GitHub Nexus connector~\cite{mcp}, 
surfacing enterprise Git repositories to the agent pipeline 
without custom integration. The LLM layer is configurable at 
runtime between GitHub Copilot Enterprise~\cite{copilot} and 
Claude API~\cite{anthropic} via a YAML configuration file, making 
the framework model-agnostic and adaptable to organizational 
licensing constraints.

\subsection{Evaluation Setup}

We report an industrial case study of a single enterprise 
deployment. Due to confidentiality constraints, all service and 
repository identifiers are anonymized; no client or system names 
are disclosed. The programme comprises 18 modern services spanning 
39 repositories. Token and coverage measurements were captured from 
LangSmith execution traces for 14 of these services; the remaining 
four are included at the programme level for effort and cost 
comparison but were not instrumented for per-run token capture. 
Coverage figures therefore report the 14 measured services, while 
the programme-level comparison in Table~\ref{tab:comparison} spans 
the full 18-service, 39-repository scope. One legacy pilot feature 
establishes cross-repository feasibility; the legacy results are 
reported as feasibility evidence, not as a general claim. We evaluate against three research questions:
\textbf{RQ1:} Can a multi-agent LLM framework systematically 
recover structured business requirements from undocumented 
enterprise software systems, both legacy and modern, with 
sufficient coverage and fidelity for downstream SDLC use? \textbf{RQ2:} Does agentic requirements recovery achieve 
meaningful reduction in effort and cost compared to manual reverse 
engineering, without sacrificing artifact accuracy?
\textbf{RQ3:} How faithfully do agentically generated bRDs capture 
system behavior, as assessed against existing test coverage and 
independent defect records?

\noindent\textbf{(a) RQ1: Coverage and Artifact Scale.} Table~\ref{tab:coverage} reports the coverage footprint across 
14 measured services from LangSmith traces. The pipeline 
discovered 280 business journeys, extracted 1,612 business rules, 
and mapped 327 test cases to generated artifacts, drawing evidence 
simultaneously from source code, test suites, configuration files, 
and documentation. This multi-source integration directly 
addresses the first limitation identified in 
Section~\ref{sec:intro}: no single source type would have 
recovered this behavioral signal alone. Journey discovery ranged 
from 3 to 42 per service and rules from 10 to 187, reflecting 
genuine complexity variation across the portfolio. Average 
generation time, the automated compute time per service exclusive 
of human review and validation, was 8.78 minutes. The 
machine-readable bRD produced for each service is directly 
consumable by downstream design, development, QA, and DevOps agents 
without translation overhead, addressing the SDLC-ready design 
principle stated in Section~\ref{sec:framework}.

\begin{table}[ht]
\centering
\caption{Coverage footprint across 14 measured services.}
\label{tab:coverage}
{\small
\rowcolors{2}{rowlight}{white}
\begin{tabular}{p{3.8cm}rr}
\toprule
\rowcolor{rowdark}
\textbf{Metric} & \textbf{Total} & \textbf{Avg/Service} \\
\midrule
Business journeys        & 280     & 20.0  \\
Business rules           & 1,612   & 115.1 \\
Test cases mapped        & 327     & 23.4  \\
Journey range            & 3--42   & ---   \\
Rules range              & 10--187 & ---   \\
Tests range              & 3--42   & ---   \\
Avg generation time (min)& 122.85  & 8.78  \\
\bottomrule
\end{tabular}
}
\end{table}

\noindent\textbf{RQ1 is answered affirmatively.} The framework 
recovers structured behavioral artifacts at enterprise scale, 
integrating evidence across multiple source types simultaneously, 
at an average generation time of 8.78 minutes per service.

\noindent\textbf{(b) RQ2: Effort and Cost Reduction.} We compare agentic measured effort against a manual baseline 
modeled from actual repository metrics for the 19-repository 
sample and projected to the full programme. The baseline is a 
model, not a stopwatch measurement of manual work on these 
services; it is derived using the equations below from real class, 
method, and LOC counts. Table~\ref{tab:repos} characterizes the 
repository sample; Table~\ref{tab:comparison} presents the 
programme-level comparison.
The complexity index (CI) weights are derived from IFPUG average 
component weights~\cite{ifpug2009}: class count maps to Internal 
Logical Files and External Interface Files (combined average 
weight 17, normalized to 0.45); method count maps to External 
Inputs, Outputs, and Inquiries (combined average weight 13, 
normalized to 0.26); production LOC modulates complexity within 
components (0.17); and test LOC maps to the Value Adjustment 
Factor (0.11). The comprehension rate of 100 LOC/hour is grounded 
in empirical studies establishing that program comprehension 
accounts for 52--70\% of software maintenance 
effort~\cite{minelli2015,xia2018}. The team multiplier of 1.8 
reflects the four-role composition required for manual reverse 
engineering: developer (100\% code coverage), business analyst 
(40\% business-critical paths), QA engineer (test code), and 
domain expert (validation). The 152-hour person-month and 
analysis and synthesis add-on rates follow 
COCOMO~II~\cite{boehm2000} and ISBSG phase effort 
benchmarks~\cite{isbsg2017} respectively. Labor cost uses a 
blended rate of \$58/hr derived from BLS 2024 occupational wage 
data~\cite{bls2024} and Salary.com surveys~\cite{salarycom2025}.

For the agentic cost model, three components are reported 
separately. LLM inference cost is computed from actual LangSmith 
token counts at Claude Haiku benchmark pricing, a generalizable 
reference independent of organizational licensing. The one-time 
setup phase (framework configuration, MCP connector setup, and 
team training) is non-recurring and amortizes across all future 
runs. Human review is one hour per service. Tooling license costs 
are excluded as they vary by contract. The two sides are therefore 
not accounted symmetrically: the manual baseline is a full 
loaded-labor model, whereas the agentic figure isolates the 
recurring marginal cost. We report the comparison on this basis and 
treat the resulting reduction as indicative of order of magnitude 
rather than as a controlled measurement.

\begin{tcolorbox}[formulabox,
    title=Manual Effort Derivation Model]

\textbf{\textcircled{1} IFPUG Complexity Index}~\cite{ifpug2009}
\begin{equation*}
CI = 0.45\,n_{cls} + 0.26\,n_{mth} + 
     0.17\,n_{loc} + 0.11\,n_{tst}
\end{equation*}
{\footnotesize
$n_{(\cdot)}$: min-max normalized values within dataset.
Classes $\rightarrow$ ILF+EIF (0.45);
Methods $\rightarrow$ EI+EO+EIQ (0.26);
Prod.\ LOC $\rightarrow$ complexity modulator (0.17);
Test LOC $\rightarrow$ Value Adjustment Factor (0.11).
}

\smallskip
\textbf{\textcircled{2} Base Hours}~\cite{minelli2015,xia2018}
\begin{equation*}
BaseHrs = \frac{TotalLOC}{100} \times 1.8
\end{equation*}
{\footnotesize
100 LOC/hr: purposeful RE of enterprise Java
(comprehension: 52--70\% of maintenance 
effort~\cite{minelli2015}).
1.8$\times$: four-role team multiplier
(Dev + BA + QA + Domain Expert).
}

\smallskip
\textbf{\textcircled{3} ISBSG Phase Add-ons}~\cite{isbsg2017}
\begin{equation*}
AnalysisAdd = \max(12,\ 0.18 \times BaseHrs)
\end{equation*}
\begin{equation*}
SynthAdd = \max(8,\ 0.13 \times BaseHrs)
\end{equation*}
{\footnotesize
0.18: ISBSG Specify phase ratio (RE mean 
18\%~\cite{isbsg2017}).
0.13: ISBSG Design phase ratio for reengineering.
Floors: 12 hrs (analysis), 8 hrs (synthesis).
}

\smallskip
\textbf{\textcircled{4} Total Hours, Person-Months, Cost}~\cite{boehm2000,bls2024}
\begin{align*}
Hrs  &= BaseHrs + AnalysisAdd + SynthAdd \\
PM   &= Hrs \div 152 \\
Cost &= Hrs \times \$58
\end{align*}
{\footnotesize
152 hrs/PM: COCOMO~II standard~\cite{boehm2000}.
\$58/hr: blended rate (BLS 2024~\cite{bls2024};
Salary.com 2025~\cite{salarycom2025}).
}

\end{tcolorbox}

\vspace{4pt}

\begin{tcolorbox}[formulabox,
    title=Agentic Cost Model]

\textbf{\textcircled{1} LLM Inference Cost}
\begin{equation*}
C_{LLM} = \frac{T_{in}}{10^6}{\times}\$0.80 
         + \frac{T_{out}}{10^6}{\times}\$2.40
\end{equation*}
{\footnotesize
$T_{in}$, $T_{out}$: input/output tokens from LangSmith 
traces; divided by $10^6$ per standard API pricing 
convention. Pricing: Claude Haiku benchmark.
}

\smallskip
\textbf{\textcircled{2} One-Time Setup (non-recurring)}
{\footnotesize
Three-week framework configuration, MCP connector setup, 
and team training. Incurred once; amortizes across all 
future runs. Not included in recurring cost per service.
}

\smallskip
\textbf{\textcircled{3} Human Review}
\begin{equation*}
C_{review} = 18\,hrs \times \$58 = \$1{,}044
\end{equation*}
{\footnotesize
One hour per service for output review and validation.
}

\smallskip
\textbf{Recurring LLM cost per service:}
\begin{equation*}
C_{marginal} = C_{LLM} \approx \$0.07
\end{equation*}
{\footnotesize
Based on average 47,694 tokens per service at Haiku 
benchmark pricing. Human review and setup costs excluded; 
see Table~\ref{tab:comparison}.
}
\end{tcolorbox}

\begin{table}[ht]
\centering
\caption{Repository sample characteristics 
(19 of 39 repositories; identifiers anonymized).}
\label{tab:repos}
{\small
\rowcolors{2}{rowlight}{white}
\begin{tabular}{p{3.5cm}rr}
\toprule
\rowcolor{rowdark}
\textbf{Metric} & \textbf{Min} & \textbf{Max} \\
\midrule
Production LOC      & 556     & 107,810 \\
Test LOC            & 0       & 82,378  \\
Class count         & 22      & 1,791   \\
Method count        & 47      & 15,228  \\
IFPUG CI            & 0.001   & 0.957   \\
Manual hours (repo) & 48      & 3,896   \\
\bottomrule
\end{tabular}
}
\end{table}

\begin{table}[ht]
\centering
\caption{Agentic token measurements across 14 services 
(LangSmith traces).}
\label{tab:tokens}
{\small
\rowcolors{2}{rowlight}{white}
\begin{tabular}{p{3.2cm}rrr}
\toprule
\rowcolor{rowdark}
\textbf{Metric} & \textbf{Min} & \textbf{Max} & 
\textbf{Avg/Svc} \\
\midrule
Input tokens       & 5,500  & 45,230 & 29,482 \\
Output tokens      & 3,300  & 28,150 & 18,211 \\
Total tokens       & 8,800  & 73,380 & 47,694 \\
Cost -- Haiku (\$) & 0.014  & 0.104  & 0.066  \\
Generation (min)   & 3.00   & 12.45  & 8.78   \\
\bottomrule
\end{tabular}
}
\end{table}

\begin{table*}[t]
\centering
\caption{Programme-level comparison: manual derived 
vs.\ agentic measured (39-repository programme scope).}
\label{tab:comparison}
{\small
\rowcolors{2}{rowlight}{white}
\begin{tabular}{p{4.5cm}p{4.0cm}p{4.0cm}}
\toprule
\rowcolor{rowdark}
\textbf{Metric} & 
\textbf{Manual (derived)} & 
\textbf{Agentic (measured)} \\
\midrule
Scope & 
39 repositories & 
18 services, 39 repos automated \\
Total human hours & 
34,981 hrs & 
$\sim$18 hrs review \\
Person-months & 
230 PM & 
$<$1 PM recurring \\
One-time setup & 
N/A & 
$\sim$3 months (non-recurring) \\
Team of 5 duration & 
$\sim$40 months & 
$<$1 week \\
LLM inference cost & 
N/A & 
\$1.18 (Haiku benchmark) \\
Total cost & 
\$2,028,869 & 
One-time setup $+$ \$1.18 LLM \\
Cost reduction & 
\multicolumn{2}{c}{\textbf{$>$98\%}} \\
Time reduction & 
\multicolumn{2}{c}{\textbf{99.4\%}} \\
\bottomrule
\end{tabular}
}
\end{table*}

The cost reduction exceeds 98\% even when the one-time setup 
investment is included, since setup is non-recurring and amortizes 
across all future runs. The recurring LLM inference cost per 
service is approximately \$0.07, so the marginal cost of each 
additional service is negligible relative to the modeled manual 
baseline.

\noindent\textbf{RQ2 is answered affirmatively.} On this modeled 
comparison, agentic requirements recovery reduces cost by more than 
98\% and generation time by 99.4\% relative to the manual baseline, 
while producing the coverage-rich artifacts demonstrated under RQ1.

\noindent\textbf{(c) RQ3: Artifact Quality Validation.} RQ3 asks how faithfully generated bRDs capture system behavior. Two 
checks are applied, and we are explicit about what each does and 
does not establish. Test coverage mapping establishes 
\emph{traceability} between generated journeys and existing tests; 
defect corroboration provides \emph{independent} check that 
extracted rules are faithful. 

\textit{Test coverage mapping (traceability).} For each measured 
service, generated journeys were matched to existing regression 
test cases automatically, through shared source references rather 
than manual inspection. A journey is traceable when at least one 
regression test maps to it. Across 14 services, 327 test cases were 
mapped to generated journeys. Because Agent~2 also uses test suites 
as a discovery source (Section~\ref{sec:framework}), this mapping 
demonstrates traceability and internal consistency rather than 
independent validation; the independent check is provided next.

\textit{Defect corroboration (independent).} To test faithfulness 
against a source the pipeline never saw, we used production defect 
records. When a service is tested, defects are logged and triaged, 
and developers routinely record in the triage notes the business 
rule that was violated. These triage descriptions are written by 
engineers for a different purpose and were not part of the 
pipeline's input, so they serve as an independent, human-authored 
statement of the system's real rules. For each sampled service we 
drew a set of logged defects and, using an LLM-as-judge, compared 
the rule cited in each defect's triage against the rules the 
pipeline had extracted. A defect corroborates when the bRD contains 
a rule matching the one the developer independently referenced. 
This checks the property that matters for artifact quality: that 
the agents recovered rules genuinely present in the system rather 
than generating fluent but ungrounded text. Across five sampled 
services, over 100 defects were examined; because a single triage 
may cite more than one rule, matches are counted at the rule level, 
and 115 distinct bRD rules were each corroborated by at least one 
defect triage that named the same rule. Where the judge could not 
match a defect, typically due to identifier formatting differences 
or the defect not concerning a business rule, the case was 
escalated to human review and non-rule defects were set aside as 
out of scope. Per-service defect counts and identifiers are 
withheld for confidentiality.

\noindent\textbf{RQ3 is answered affirmatively.} Extracted rules 
match the rules developers independently cited when triaging real 
production defects, evidence that the pipeline recovers faithful 
system behavior rather than plausible but ungrounded text, while 
test mapping confirms generated journeys trace to verified tests.

\section{Discussion}
\label{sec:discussion}

The framework shows that automated requirements reverse engineering 
at enterprise scale is achievable without sacrificing artifact 
quality, making bRD generation a continuous activity throughout the 
migration lifecycle rather than a deferred one. The significance is 
concrete: a comparable prior migration to the same target 
architecture for another product line required over two years of 
effort in practice, whereas the framework produces the foundational 
artifacts for an equivalent programme in days. On the modern side, 
generated bRDs give feature development and agile delivery the 
behavioral baseline they rarely have, and the defect corroboration 
shows extracted rules match those developers independently relied 
on, so the artifacts are faithful, not merely comprehensive.

Several limitations bound these results. The manual baseline is 
modeled from actual repository metrics rather than observed 
directly, and the two cost sides are not accounted symmetrically, so 
the reported reduction is an order-of-magnitude indication, not a 
controlled measurement. The study covers one enterprise Java 
codebase and a single legacy pilot; generalization to other 
languages and domains remains future work. Because extraction is 
static-first, behavior resolved only at runtime (dependency 
injection, reflection, dynamic proxies) and rules outside 
application code (stored procedures, triggers, message routing) are 
under-represented or out of scope, and coverage metrics do not 
verify completeness. Where source is sensitive, the model-agnostic 
design permits locally hosted models. The deployment is deliberately human-supervised at this maturity 
stage. Where a component lacks a usable test suite, the pipeline 
still runs, capping affected findings at lower confidence and 
reporting the gap as a risk. The planned next step is autonomous 
scheduled execution in which bRDs regenerate as code changes, using 
the knowledge layer's incremental mode to reprocess only the delta, 
countering drift between a living codebase and its documentation. 
The same principles extend to downstream agents for code generation, 
review, and testing, completing the agentic SDLC pipeline this work 
initiates.

\section{Conclusion}
\label{sec:conclusion}

Enterprise software systems carry decades of behavioral knowledge 
that exists nowhere in formal documentation. This paper presented 
an agentic framework that recovers it systematically through 
multi-source integration, static-first extraction, and embedded 
expert practitioner cognitive models. Deployed on a real enterprise 
environment spanning modern microservices and a legacy migration 
pilot, the framework shows that requirements reverse engineering is 
tractable at scale and at negligible recurring cost, with extracted 
rules corroborated against independent production defect records. 
The result is not merely a faster path to bRD generation but a 
foundation for a fully agentic SDLC pipeline, in which design, 
development, testing, and deployment agents build on a shared, 
machine-readable specification the knowledge layer maintains as the 
codebase evolves.

% if added before the last page, this command can help balancing columns
%\addtolength{\textheight}{-.2cm} 

%Bibliography 
% \bibliographystyle{apalike}
% \bibliography{sample}

\printbibliography

\end{document}